\documentclass[fleqn]{article}

\usepackage{amssymb,amsmath}
\usepackage{dcolumn}
\newcolumntype{.}{D{.}{.}{-1}}
\newcolumntype{d}{D{.}{.}{-1}}
\usepackage{graphicx}
\usepackage{authblk}
\graphicspath{ {C:/Users/Sayan/Desktop/Publication fundamental constants} }

\newcommand{\HDone}[1]{%
    $v:0\rightarrow 1$ }
\newcommand{\HDfour}[1]{%
    $v:0\rightarrow 4$ }
\newcommand{\HDeight}[1]{%
    $v:0\rightarrow 8$ }

\begin{document}

\title{Proton-electron mass ratio from HD$^+$ revisited}
\date{\vspace{-5ex}}
\author[1]{\small Sayan~Patra}
\author[2,3]{\small J.-Ph.~Karr}
\author[2,3]{\small L.~Hilico}
\author[1]{\small M.~Germann}
\author[4]{\small V.~I.~Korobov}
\author[1]{\small J.~C.~J.~Koelemeij}

\affil[1]{\footnotesize LaserLaB and Department of Physics and Astronomy, Vrije Universiteit Amsterdam\\
De Boelelaan 1081, 1081 HV Amsterdam, The Netherlands}
\affil[2]{\footnotesize Laboratoire Kastler Brossel, UPMC-Sorbonne Universit\'es, CNRS, ENS-PSL Research University, Coll\`ege de France\\
4 Place Jussieu, F-75005 Paris, France}
\affil[3]{\footnotesize D\'epartement de Physique, Universit\'e d'Evry-Val d'Essonne, Boulevard Fran\c cois Mitterrand, F-91000 Evry, France}
\affil[4]{\footnotesize Bogoliubov Laboratory of Theoretical Physics, Joint Institute for Nuclear Research, Dubna 141980, Russia}
\maketitle

\begin{abstract}
We present a new derivation of the proton-electron mass ratio from the hydrogen molecular ion, HD$^+$. The derivation entails the adjustment of the mass ratio in highly precise theory so as to reproduce accurately measured ro-vibrational frequencies. This work is motivated by recent improvements of the theory, as well as the more accurate value of the electron mass in the recently published CODATA-14 set of fundamental constants, which justifies using it as input data in the adjustment, rather than the proton mass value as done in previous works. This leads to significantly different sensitivity coefficients and, consequently, a different value and larger uncertainty margin of the proton-electron mass ratio as obtained from HD$^+$.
\end{abstract}

\section{Introduction}

Hydrogen molecular ions, on account of their simple three-body structure, are the simplest molecules in nature. Because of this, they are benchmark systems for testing molecular theory. \textit{Ab initio} calculations of the ro-vibrational transition frequencies in the ground electronic state of these molecules can be done with very high accuracy. Recently, the fundamental ro-vibrational transition frequencies of H$_2^+$ and HD$^+$ were calculated with relative uncertainties of about $8\times10^{-12}$~\cite{KorobovPRL2017}. These calculations include relativistic, radiative (QED) and nuclear finite-size corrections to the non-relativistic energies of the ro-vibrational levels. An experiment performed with similar or better accuracy than the theoretical predictions would not only allow a stringent test of the calculations, but also of the theoretical framework within which the calculations were done. Moreover, as suggested by Wing \textit{et al.} more than four decades ago, such an experiment might lead to an improved determination of several fundamental constants, in particular the proton-electron mass ratio, $\mu_{\rm pe}$~\cite{Wing1976}. On the experimental front, with the advances in charged particle trapping combined with laser-cooling techniques and high-resolution laser spectroscopy, progress has been made towards achieving an accuracy comparable to (and ultimately better than) the theoretical predictions. In 2007, the $(v,L):(0,2)\rightarrow(4,3)$ overtone in HD$^+$ was measured with a relative frequency uncertainty of 2 parts-per-billion (ppb)~\cite{Koelemeij2007}. In 2012, the fundamental transition $(v,L):(0,0)\rightarrow(1,1)$ in HD$^+$ was measured with a relative uncertainty of 1.1~ppb~\cite{Bressel2012}. In this case, the experimentally determined `spin-averaged' ro-vibrational transition frequency was found to be offset from the more accurate theoretical prediction by 2.5~$\sigma$. More recently, a measurement of the $(v,L):(0,2)\rightarrow(8,3)$ transition in the same molecule with 1.1~ppb relative uncertainty agreed with the theoretical predictions within the combined experimental and theoretical uncertainty~\cite{BiesheuvelNatComm2016,BiesheuvelAPB2017}. Because of the agreement between the theoretical prediction and the experimental determination, the authors could for the first time extract the value of  $\mu_{\rm pe}$ as a single parameter with a relative uncertainty of 2.9 ppb~\cite{BiesheuvelNatComm2016}.

In this article, we revisit the determination of $\mu_{\rm pe}$ from HD$^+$ taking into account all existing measurements of ro-vibrational transitions in HD$^+$, similar as done by Karshenboim and Ivanov~\cite{KarshenboimAPB2017}. Since the publication of Ref.~\cite{BiesheuvelNatComm2016,BiesheuvelAPB2017,KarshenboimAPB2017}, theoretical calculations were improved by including previously unaccounted higher-order QED correction terms~\cite{KorobovPRL2017}. The improvement in theoretical calculations by itself forms a reason to re-enumerate its agreement with the previously measured transitions. Also, since the improved calculations were performed using newly published CODATA-14 recommended values of the fundamental constants, in this article we present a consistent determination of $\mu_{\rm pe}$ from HD$^+$. As we will explain further below, the strongly improved value of the new CODATA-14 value of the electron mass (in atomic mass units) makes it more appropriate to include it as input data instead of the proton mass value, which affects the sensitivity coefficients used in previous determinations of $\mu_{\rm pe}$ from HD$^+$. We subsequently present a revised value of $\mu_{\rm pe}$ here.

This article is structured as follows: In section \ref{sec:II}, we briefly review the recent improvements in the theoretical calculation of ro-vibrational transitions in the ground electronic state of HD$^+$, followed by a comparison between the existing measurements with the improved theory in section \ref{sec:III}. In section \ref{sec:IV}, we determine the proton-to-electron mass ratio $\mu_{\rm pe}$ from the measurements considered in section \ref{sec:III}. In section \ref{sec:V}, we discuss the prospects of an improved determination of $\mu_{\rm pe}$ from Doppler-free two-photon spectroscopy of HD$^+$ and H$_2^+$.

\section{Improvement in theory}\label{sec:II}
The energy of a ro-vibrational level of HD$^+$, calculated in the framework of QED, may be written as
\begin{equation}
\label{energy}
E = R_{\infty}\!\left[ E_{\rm nr} (\mu_{\rm pe},\mu_{\rm de}) \!+ \alpha^2 F_{\rm QED} (\alpha) + A^{\rm fs}_{\rm p} \left( \frac{r_{\rm p}}{a_{\rm 0}} \right)^2 + A^{\rm fs}_{\rm d} \left( \frac{r_{\rm d}}{a_{\rm 0}} \right)^2 \right]
\end{equation}
where $R_{\infty}$ and $\alpha$ are the Rydberg and fine-structure constant respectively, and $a_{\rm 0} = \alpha/4\pi R_{\infty}$ is the Bohr radius. The main contribution to $E$ is the non-relativistic (Schr\"odinger) energy $E_{\rm nr}$, which depends on the proton-electron ($\mu_{\rm pe}$) and deuteron-electron ($\mu_{\rm de}$) mass ratios. The next term corresponds to relativistic and QED corrections. The function $F_{\rm QED} (\alpha)$ is a non-analytic expansion which, beyond powers of $\alpha$, also contains logarithmic terms like $\alpha^p \ln^q(\alpha)$. The last two terms are the leading-order nuclear finite-size corrections, with $r_{\rm p}$ and $r_{\rm d}$ respectively standing for the proton and deuteron charge radii. The coefficients $A^{\rm fs}_{\rm p,d}$ are proportional to the squared density of the wave function at the electron-nucleus coalescence point. Higher-order nuclear size and structure corrections are negligible at the current level of theoretical accuracy.

The non-relativistic energy $E_{\rm nr}$ and its dependence on the mass ratios can be calculated with very high accuracy by numerical solution of the Schr\"odinger equation for the exact three-body Coulomb Hamiltonian using a variational method (see e.g.~\cite{Schil05,Karr06,Li07,Ning14}). The theoretical accuracy of the energy levels $E$ has been steadily improved over the last decade through a systematic evaluation, in ascending powers of $\alpha$, of the QED contributions appearing in $F_{\rm QED} (\alpha)$ in the framework of non-relativistic QED (NRQED). The main steps of this work have been published in the successive papers~\cite{KorobovPRL2017,Kor06,Kor08,Kor14PRA,Kor14PRL}. The first step~\cite{Kor06} was the calculation of leading relativistic and radiative corrections at the $R_{\infty} \alpha^2$ and $R_{\infty} \alpha^3$ orders within an exact three-body approach, with a  partial consideration of contributions at the following order ($R_{\infty} \alpha^4$). This was later pursued~\cite{Kor08} by a calculation of relativistic corrections at this order in the framework of the adiabatic approximation. The first high-precision comparisons between theory and experiment involving the ro-vibrational spectrum of HD$^+$~\cite{Koelemeij2007,Bressel2012} were done with the predictions from~\cite{Kor06,Kor08}.

A few years later, the theory was further refined by the calculation of $R_{\infty} \alpha^5$-order corrections~\cite{Kor14PRA,Kor14PRL} within the adiabatic approach. These results were used in the analysis of the recently measured $(v,L)=(0,2) \to (8,3)$ transition~\cite{BiesheuvelNatComm2016,BiesheuvelAPB2017}.

However, it has since been realized~\cite{Kor16} that the treatment of second-order perturbation terms in the adiabatic approximation as done in these previous works was incomplete. This type of contribution is present in the $R_{\infty} \alpha^4$-order relativistic correction, and also appears at higher orders like in the $R_{\infty} \alpha^5$-order one-loop corrections. Since this gives the largest contribution to the difference between earlier works~\cite{Kor06,Kor08,Kor14PRA,Kor14PRL} and the updated predictions used here~\cite{KorobovPRL2017}, it is worth explaining this point in more detail.

The general structure of such terms is
\begin{equation}
\Delta E = \left\langle \Psi \right| A Q (E_0-H)^{-1} Q B \left| \Psi \right\rangle + \left\langle \Psi \right| H^{(n)} \left| \Psi \right\rangle \label{order2}
\end{equation}
where $A$, $B$ and $H^{(n)}$ are effective operators acting on the electron, $Q$ is a projection operator onto a subspace orthogonal to the non-relativistic wavefunction $\Psi$, $E_0,H$ the non-relativistic energy and Hamiltonian. For example, a term of the type~(\ref{order2}) with $A = B = H_B$, where $H_B$ is the electronic Breit-Pauli Hamiltonian, appears in the $R_{\infty} \alpha^4$ relativistic correction~\cite{Kor08}. Similarly, a term with $A = H_B$ and $B = U_{\rm vp}$ (and an additional factor of 2), where $U_{\rm vp}$ is the electron-nuclei Uehling interaction, appears in the one-loop vacuum polarization at the $R_{\infty} \alpha^5$ order~(see Eq.~(4) of~\cite{Karr17}). In the adiabatic approximation, the molecular wave function is taken in a form
\begin{equation}
\Psi(\mathbf{r},R) = \phi_{\rm el}(\mathbf{r};R) \chi_{\rm ad}(R)
\end{equation}
where $\phi_{\rm el}(\mathbf{r};R)$ and $\chi_{\rm ad}(R)$ are respectively the electronic and nuclear wave functions. Then Eq.~(\ref{order2}) can be written as a sum over intermediate states which may be separated into three terms. The first term involves only electronic excitations:
\begin{subequations}
\begin{equation}
\Delta E_{\rm el} = \left\langle \chi_{\rm ad} \right| \Delta \mathcal{E}_{\rm el} (R) \left| \chi_{\rm ad} \right\rangle \label{elec},
\end{equation}
\begin{equation}
\Delta \mathcal{E}_{\rm el} (R) = \left\langle \phi_{\rm el} \right| A Q_{\rm el} (E_{\rm el}-H_{\rm el})^{-1} Q_{\rm el} B \left| \phi_{\rm el} \right\rangle + \left\langle \phi_{\rm el} \right| H^{(n)} \left| \phi_{\rm el} \right\rangle.
\end{equation}
\end{subequations}
Here $Q_{\rm el}$ is a projection operator onto a subspace orthogonal to $\phi_{\rm el}$, and $E_{\rm el}, H_{\rm el}$ the electronic energy and Hamiltonian. The second term involves vibrational excitations:
\begin{equation}
\Delta E_{\rm vb} = \left\langle \chi_{\rm ad} \right| \mathcal{A}(R) Q_{\rm vb} (E_{\rm vb}-H_{\rm vb})^{-1} Q_{\rm vb} \mathcal{B}(R) \left| \chi_{\rm ad} \right\rangle \label{vibr},
\end{equation}
with $\mathcal{A}(R) = \langle \phi_{\rm el} | A | \phi_{\rm el} \rangle$, $\mathcal{B}(R) = \langle \phi_{\rm el} | B | \phi_{\rm el} \rangle$, $Q_{\rm vb}$ is a projection operator onto a subspace orthogonal to $\chi_{\rm ad}$, and $E_{\rm vb}, H_{\rm vb}$ the vibrational energy and Hamiltonian. Finally, the third term is the contribution beyond the adiabatic approximation, involving simultaneous electronic and vibrational excitations. This term is very small and may be neglected at the current level of theoretical accuracy, as has been explicitly verified in the case of the $R_{\infty} \alpha^5$-order one loop vacuum polarization contribution~\cite{Karr17} by comparing the sum of Eqs.~(\ref{elec}) and~(\ref{vibr}) to a full calculation of Eq.~(\ref{order2}) performed in an exact three-body approach.

The vibrational contribution of Eq.~(\ref{vibr}) is the term which had been neglected in previous treatments. Although its value for individual ro-vibrational states is significantly smaller than the respective electronic contribution of Eq.~(\ref{elec}) (by typically one order of magnitude), its contribution to ro-vibrational transition frequencies is more important because it has a stronger dependence on the ro-vibrational state leading to a much less pronounced cancellation. For example, the vibrational part of the $R_{\infty} \alpha^4$ relativistic correction contributes to the transition frequencies at a relative level of about $7 \times 10^{-10}$, which is comparable to the experimental uncertainties.

In addition to a systematic evaluation of vibrational terms at the $R_{\infty} \alpha^4$ and $R_{\infty} \alpha^5$ orders, the work of Ref.~\cite{KorobovPRL2017} improved the theoretical accuracy further through a partial calculation of the following order $R_{\infty} \alpha^6$. It has also proved necessary to improve the numerical accuracy of the leading-order relativistic and radiative corrections (especially the Bethe logarithm)~\cite{Zhong12,Kor12}, since their initial evaluation~\cite{Kor06} targeted a lower theoretical precision. Updated theoretical predictions for the three most accurately measured transitions in HD$^+$ are given in Table~\ref{theor}. The relative theoretical uncertainty is about $8 \times 10^{-12}$ in all cases.

\begin{table*}[t]
\begin{center}
\caption{Theoretical transition frequencies for the three most accurately measured ro-vibrational transitions in HD$^+$ (in kHz). The first line is the non-relativistic transition frequency, and next are QED corrections in ascending powers of $\alpha$. The final result given in the last line is the sum of all the above terms with an additional (very small) muonic vacuum polarization contribution. Nuclear finite-size corrections have been included in $\Delta E_{\alpha^2}$ for simplicity. The CODATA-14 recommended values of fundamental constants are used. Estimated theoretical uncertainties, when significant, are given between parentheses.}
\label{theor}
\begin{tabular}{cddd}
\hline\hline
\vrule height 10.5pt width 0pt depth 3.5pt
\text{Transition} & \multicolumn{1}{c}{$(0,0) \to (1,1)$} & \multicolumn{1}{c}{$(0,2) \to (4,3)$} & \multicolumn{1}{c}{$(0,2) \to (8,3)$} \\
\hline
\vrule height 10pt width 0pt
$\Delta E_{\rm nr}$ & 58\,604\,301\,246.9 & 214\,976\,047\,255.7 & 383\,403\,254\,198.4     \\
$\Delta E_{\alpha^2}$ & 1\,003\,551.5 & 3\,411\,243.9 & 5\,470\,087.2     \\
$\Delta E_{\alpha^3}$ & -250\,978.4 & -891\,610.9(3) & -1\,536\,834.7(5)  \\
$\Delta E_{\alpha^4}$ & -1\,770.8 & -6\,307.9(1) & -10\,914.3(1)  \\
$\Delta E_{\alpha^5}$ & 110.3 & 352.8(1) & 684.1(2)  \\
$\Delta E_{\alpha^6}$ & -2.1(5) & -7.6(17) & -13.7(29) \\
\hline
\vrule height 10pt width 0pt
$\Delta E_{\rm tot}$& 58\,605\,052\,157.5(5) & 214\,978\,560\,967.8(17) & 383\,407\,177\,208.0(30) \\
\hline\hline
\end{tabular}
\end{center}
\end{table*}

\section{Comparison between experiment and theory}\label{sec:III}

Experimentally measured `spin-averaged' frequencies of the different ro-vibrational transitions in HD$^+$ are tabulated along with their respective experimental uncertainties in Table~\ref{theoexptcomp}, together with the corresponding theoretical transition frequencies.
\begin{table}[h!]
\begin{center}
\caption{Comparison of the frequencies of the three most accurately measured ro-vibrational transitions in HD$^+$ with their corresponding theoretical predictions. The second column presents the transition frequencies calculated from first principles, while the third column presents the corresponding measured transition frequencies. The uncertainties are shown in parentheses. The last column gives the values of the sensitivity coefficient to $\mu_{\rm pe}$, which is defined and discussed in Section~\ref{sec:IV}.}
\label{theoexptcomp}
\begin{tabular}{c . . .}
 \hline
 \hline
 \multicolumn{1}{c}{Transition} & \multicolumn{1}{c}{$\nu_{\rm theo}$ [MHz]} & \multicolumn{1}{c}{$\nu_{\rm exp}$ [MHz]} & \multicolumn{1}{c}{$S_{\rm pe(de)}^{fi}$} \\
  \hline
   \ {$(v,L): (0,0)\rightarrow(1,1)$} & 58\,605\,052.1575 (5) & 58\,605\,052.000 (64) & -0.32296 \\
   \ {$(v,L): (0,2)\rightarrow(4,3)$} & 214\,978\,560.9678 (17) & 214\,978\,560.6 (5) & -0.29190 \\
   \ {$(v,L): (0,2)\rightarrow(8,3)$} & 383\,407\,177.208 (3) & 383\,407\,177.38 (41) & -0.24998 \\
 \hline
 \hline
\end{tabular}
\end{center}
\end{table}

\begin{figure}[!ht]
\begin{center}
\includegraphics[scale=0.6]{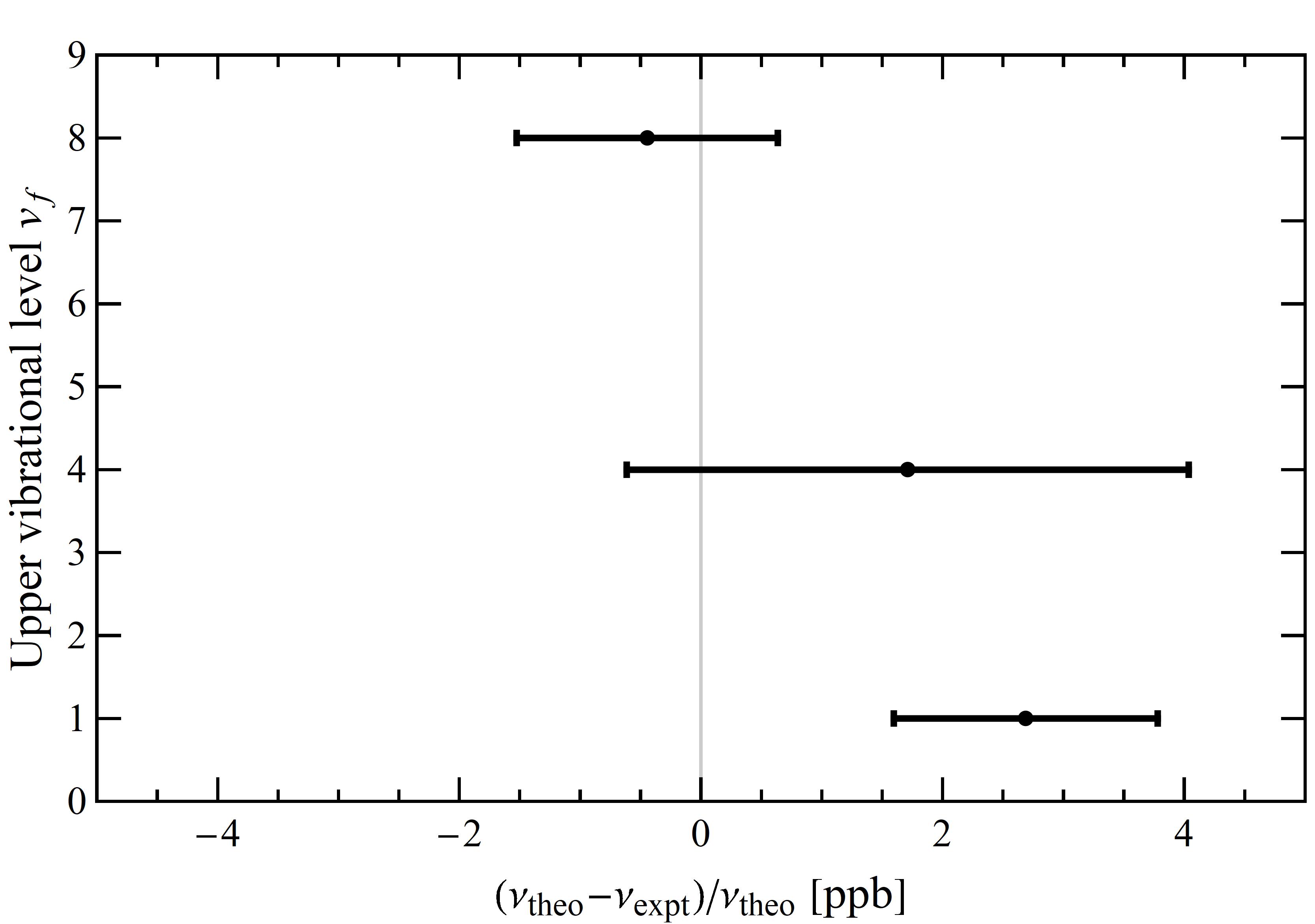}
\caption{Relative offsets of the measured frequencies of the three ro-vibrational transitions in HD$^+$ with respect to their respective theoretical predictions. Error bars represent the combined theoretical and experimental uncertainty, $\sigma_c$.}
\label{fig:expvstheo}
\end{center}
\end{figure}

Figure~\ref{fig:expvstheo} graphically shows the level of agreement between the measured transition frequencies and their respective theoretical predictions. Here it should be noted that the error bars represent the combined uncertainty of experiment and theory, given by $\sigma_c =(\sigma_e^2+\sigma_t^2)^{1/2}$. It can be observed that the $(v,L):(0,2)\rightarrow(4,3)$ measurement (henceforth indicated by the shorthand \HDfour\ , and with similar notations for the other two transitions) has the largest relative (combined) uncertainty of 2.3~ppb, while the offset from the theoretical prediction is 1.7~ppb. The transitions \HDone\  and \HDeight\  have similar experimental uncertainties of 1.1 ppb. However, the former deviates from theory by 2.7 ppb (2.5~$\sigma$). The experiments are described in detail elsewhere~\cite{Koelemeij2007,BiesheuvelAPB2017,Bressel2012}.

\section{Determination of the proton-electron mass ratio $\mu_{\rm pe}$}\label{sec:IV}

From Eq.~(\ref{energy}) it may be observed that the ro-vibrational transition frequencies in HD$^+$ depend on no less than six fundamental constants: the Rydberg constant $R_{\infty}$, the fine structure constant $\alpha$, the nuclear radii $r_{\rm p},r_{\rm d}$ and the nucleus-to-electron mass ratios $\mu_{\rm pe},\mu_{\rm de}$. However, four of them ($R_{\infty}$, $\alpha$, $r_{\rm p}$ and $r_{\rm d}$) have been determined by other experiments with an accuracy that is several order of magnitude higher than could be obtained from HD$^+$ spectroscopy experiments performed so far. We thus fix their values as equal to the CODATA recommended ones and focus on determination of nucleus-to-electron mass ratios.

It may seem that the three measurements compiled in Table~\ref{theoexptcomp} are sufficient to simultaneously constrain the two independent parameters $\mu_{\rm pe}$ and $\mu_{\rm de}$. However, it is important to realize that all the HD$^+$ ro-vibrational spacings essentially depend on the ratio $\mu_{\rm re}$ of the nuclear reduced mass $m_{\rm r} = m_{\rm p} m_{\rm d}/(m_{\rm p}+m_{\rm d})$ to the electron mass. Consequently, measurements in HD$^+$ only allow determining $\mu_{\rm re}$, but not $\mu_{\rm pe}$ and $\mu_{\rm de}$ separately (see also the discussion in Ref.~\cite{Schil05}). This issue could potentially be solved in the future by combining the HD$^+$ measurements with the measurement of one or several transitions in H$_2^+$, whose frequencies only depend on $\mu_{\rm pe}$~\cite{KarrPRA2016}. Here, having only HD$^+$ data at our disposal, we need to set one of the mass ratios to its CODATA value in order to determine another one.

Note that the choice of variables $(\mu_{\rm pe},\mu_{\rm de})$ made in Eq.~(\ref{energy}) is arbitrary: in principle, we could fix any one of the three mass ratios $(\mu_{\rm pe},\mu_{\rm de},\mu_{\rm dp})$. The most relevant choice depends on the current state of knowledge, since we should set the value of the most accurately known mass ratio in order get a new determination of another, less accurately known one. In this regard, an important evolution has occurred in the latest adjustment of fundamental constants (CODATA-14). In the CODATA-10 set of recommended values, the electron mass (in atomic mass units) had a relative uncertainty of $4.0\times10^{-10}$, while the relative uncertainties in the values of $m_{\rm p}$ and $m_{\rm d}$ were $8.9\times10^{-11}$ and $3.8\times10^{-11}$ respectively~\cite{CODATA10}. Thus the most accurately known mass ratio was $\mu_{\rm dp}$; this has led to fix $\mu_{\rm dp}$ in previous adjustments of $\mu_{\rm pe}$. However, with the publication of the CODATA-14 recommended values, the relative uncertainty of $m_{\rm e}$ was reduced to $2.9\times10^{-11}$, while the relative uncertainties in $m_{\rm p}$ and $m_{\rm d}$ were reported to be $9.0\times10^{-11}$ and $2.0\times10^{-11}$ respectively~\cite{CODATA14}. The most accurately known mass ratio is now $\mu_{\rm de}$, so that it makes more sense to set it to its CODATA-14 value in adjusting $\mu_{\rm pe}$. We will follow this new approach in the present work.

For a deeper understanding of the reasons for this choice, it is useful to observe that the CODATA values of the mass ratios are essentially obtained from separate determinations of the particle masses, by mass spectrometry ($m_{\rm p}$, $m_{\rm d}$) or $g$-factor measurements ($m_{\rm e}$). Fixing the value of $\mu_{\rm dp}$, as done in previous treatments, is equivalent to taking into account the measurements of $m_{\rm d}$ and $m_{\rm p}$, while ignoring that of $m_{\rm e}$. In this case, the value of $\mu_{\rm pe}$ extracted from HD$^+$ spectroscopy may be interpreted as a cross-check of the electron mass measurement. Here, we will fix $\mu_{\rm de}$, meaning that we take into account the measurements of $m_{\rm d}$ and $m_{\rm e}$, but ignore that of $m_{\rm p}$. Our subsequent determination of $\mu_{\rm pe}$ can be seen as a consistency check of the proton mass value, as obtained from mass spectrometry and from molecular spectroscopy. In the present state of knowledge, it is more relevant to cross-check the proton mass than the electron mass since the latter has been determined to higher accuracy.

In order to derive the proton-electron mass ratio from the data of Table~\ref{theoexptcomp}, we need to calculate the dependence of the ro-vibrational transition frequencies on $\mu_{\rm pe}$. This
can be expressed in terms of a normalized sensitivity coefficient, $S_{\rm pe(\textit{n})}^{fi}$, as
\begin{eqnarray}
S_{\rm pe(\textit{n})}^{fi} &=& \frac{\mu_{\rm pe}}{\nu_{fi}} \left. \frac{\partial \nu_{fi}}{\partial \mu_{\rm pe}} \right|_{\mu_n} \\
                   &=& \frac{1}{(E_f-E_i)} \left( \mu_{\rm pe} \left. \frac{\partial E_f}{\partial \mu_{\rm pe}} \right|_{\mu_n} - \mu_{\rm pe} \left. \frac{\partial E_i}{\partial \mu_{\rm pe}}\right|_{\mu_n} \right). \label{sensit-trans}
\label{sensitivity}
\end{eqnarray}
Here $\nu_{fi}$ is the frequency of a transition from a lower level $i$ with energy $E_i$ to an upper level $f$ with energy $E_f$, and $\mu_n$ ($n = {\rm dp}$ or ${\rm de}$) denotes the mass ratio that is kept fixed while varying $\mu_{\rm pe}$.

The main dependence of the energy levels given by Eq.~(\ref{energy}) on mass ratios arises from the non-relativistic contribution $E_{\rm nr}$. Although the QED correction terms also depend on the mass ratios, their contribution to the overall dependence can be neglected since they are smaller by a factor of $\alpha^2$. The sensitivity coefficients $S_{\rm pe(\textit{n})}^{fi}$ can thus be obtained with sufficient accuracy from a calculation of nonrelativistic energy levels as done in~\cite{Schil05,Karr06}. These works used the variables $(\mu_{\rm pe},\mu_{\rm dp})$ and provide the coefficients $\mu_{\rm pe} \left. \frac{\partial E}{\partial \mu_{\rm pe}} \right|_{\mu_{\rm dp}}$ for individual ro-vibrational levels, from which the sensitivities $S_{\rm pe(dp)}^{fi}$ of ro-vibrational transitions were obtained by applying Eq.~(\ref{sensit-trans}) and used in previous determinations of $\mu_{\rm pe}$ from HD$^+$~\cite{BiesheuvelNatComm2016,KarshenboimAPB2017}. Since in our new approach we fix $\mu_{\rm de}$, we use a different coefficient, $S_{\rm pe(de)}^{fi}$. To obtain this coefficient, it is convenient to write the non-relativistic three-body Hamiltonian in terms of $\mu_{\rm pe}$ and $\mu_{\rm de}$, following the notations of Eq.~(6) in~\cite{Schil05}:
\begin{equation}
H_0 = -\frac{1}{2} \left( \mu_{\rm pe}^{-1} + \mu_{\rm de}^{-1} \right) \nabla_{\mathbf{r_1}}^2 - \frac{1}{2} \left( 1 + \mu_{\rm de}^{-1} \right) \nabla_{\mathbf{r_2}}^2 - \mu_{\rm de}^{-1} \; \nabla_{\mathbf{r_1}} \cdot \nabla_{\mathbf{r_2}} + V_C,
\end{equation}
where $\mathbf{r_1}$ and $\mathbf{r_2}$ are the position vectors of proton and electron with respect to the deuteron, and $V_C$ the Coulomb interaction potential. Using Eqs.~(10) of~\cite{Schil05} one immediately gets
\begin{equation}
\mu_{\rm pe} \left.  \frac{\partial E}{\partial \mu_{\rm pe}} \right|_{\mu_{\rm de}}=\frac{1}{2} \; \mu_{\rm pe}^{-1} \left\langle \nabla_{\mathbf{r_1}}^2 \right\rangle \label{new-coeff-1} \\
\end{equation}
The quantity on the LHS of Eq.~(\ref{new-coeff-1}) can also be expressed as shown below using the chain rule of differential calculus,
\begin{equation}
\mu_{\rm pe} \left.  \frac{\partial E}{\partial \mu_{\rm pe}} \right|_{\mu_{\rm de}}=\mu_{\rm pe} \left. \frac{\partial E}{\partial \mu_{\rm pe}} \right|_{\mu_{\rm dp}} - \mu_{\rm dp} \left. \frac{\partial E}{\partial \mu_{\rm dp}} \right|_{\mu_{\rm pe}} \label{new-coeff-2}
\end{equation}
The quantities appearing on the RHS of Eq.~\ref{new-coeff-2} are given in Tables~II and~III of~\cite{Schil05} for a range of ro-vibrational states. We used these values to get the sensitivity coefficients for the \HDone\ and \HDfour\ transitions. For the \HDeight\ transition we directly determined the sensitivity of the $(v,L): (8,3)$ level from Eq.~(\ref{new-coeff-1}). The values of the sensitivity coefficients for the three transitions can be found in Table~\ref{theoexptcomp}.

As discussed in~\cite{Schil05}, the underlying dependence of ro-vibrational energies on the ratio $\frac{\mu_{\rm re}}{m_{\rm e}}$ results in a fixed ratio between the various sensitivity coefficients for a given level. For example, in Eq.~(\ref{new-coeff-2}) the first term is almost exactly three times larger than the second one~\cite{Schil05}. This implies that $\left. \frac{\partial E}{\partial \mu_{\rm pe}} \right|_{\mu_{\rm de}}$ is about $2/3$ times $\left. \frac{\partial E}{\partial \mu_{\rm pe}} \right|_{\mu_{\rm dp}}$, and therefore the sensitivities of the transition frequencies used in our approach, $S_{\rm pe(de)}^{fi}$ are also smaller by a factor of $2/3$ with respect to the coefficients used in previous work, $S_{\rm pe(dp)}^{fi}$.  It then follows that the error bar and displacement of the found value of $\mu_{\rm pe}$ from the CODATA-14 value are about $3/2$ times those published by Biesheuvel \textit{et al.}~\cite{BiesheuvelNatComm2016} and Karshenboim and Ivanov~\cite{KarshenboimAPB2017}. As justified above, the value and uncertainty of $\mu_{\rm pe}$ reported in this work supersede the previously reported values.

From a single transition measurement, a new value of $\mu_{\rm pe}$ is deduced using the relation
\begin{equation}
\mu_{\rm pe}(\nu_{fi})=\mu_{\rm pe,0}+\frac{\mu_{\rm pe,0}}{S_{\rm pe(de)}^{fi}}\frac{\nu_{fi,\rm exp}-\nu_{fi,\rm theo}}{\nu_{fi,\rm theo}},
\label{eqnmu}
\end{equation}
and if we choose to combine several transition measurements, then the resulting value of $\mu_{\rm pe}$ is obtained by a standard least-squares adjustment procedure.

\begin{figure}[!ht]
\begin{center}
\includegraphics[scale=0.1]{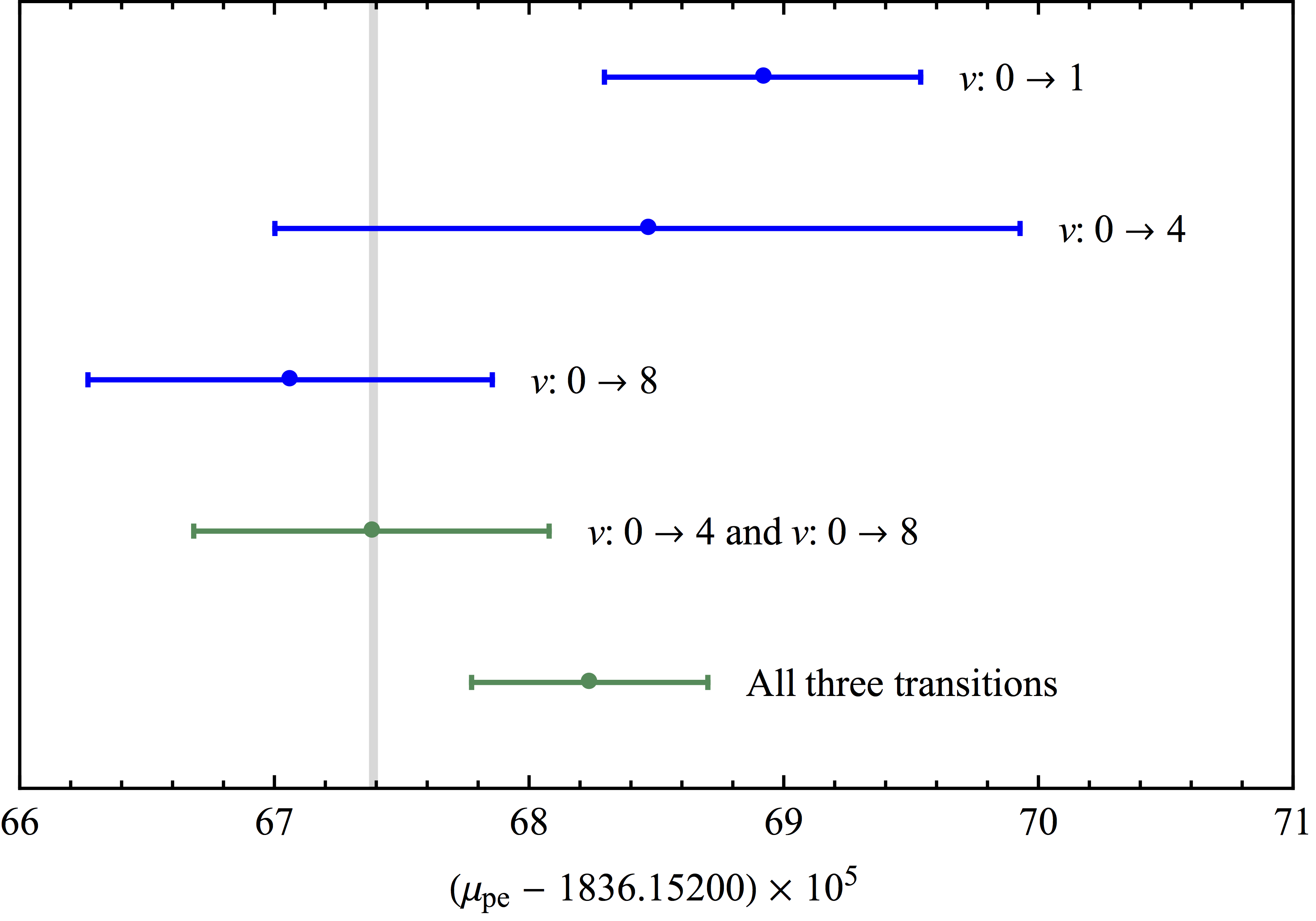}
\caption{Visual representation of the results of the various $\mu_{\rm pe}$ values determined from measurements in HD$^+$, as compared to the CODATA-14 recommended value $\mu_{\rm pe}$(C14) (value and error bar indicated by the position and width of the vertical grey line).}
\label{mucomparison}
\end{center}
\end{figure}

Before other transitions are included, it is important to consider the two key assumptions that underpin the determination of fundamental constants from the comparison of experiment and theory, as done here for $\mu_{\rm pe}$. These assumptions are that the theory predicts the observable properties of the system under consideration faithfully, and that the experimental measurement is performed without any uncorrected biases. Consequently, any significant disagreement between theory and experiment (in terms of the combined experimental and statistical uncertainty) indicates that at least one of the two key assumptions is likely not met (depending on a pre-specified minimum required confidence level). Of the three transitions in HD$^+$ considered here, the \HDfour\ and \HDeight\ measurements agree with theory within one sigma ($\sigma_c$). Of these two transitions, the \HDfour\ transition contributes relatively little given its 2.3~ppb uncertainty (to be compared with 1.1~ppb for the \HDeight\ transition). The \HDone\ transition has a small combined relative uncertainty of 1.1~ppb, but displays a (hitherto unresolved) discrepancy of $2.5~\sigma_c$ (see figure~\ref{fig:expvstheo}). Depending on the minimum confidence level required (a subject which we will not address here) this transition therefore may or may not be taken into account in the determination of $\mu_{\rm pe}$. We therefore provide results for both scenarios. In Table~\ref{muoverview}, an overview of the determination of $\mu_{\rm pe}$ from different combinations of the measurements as well as individual transitions considered is presented. A visual representation of the deviations of $\mu_{\rm pe}$ extracted from HD$^+$ from the CODATA-14 recommended value is provided in Figure~\ref{mucomparison}.
\begin{table}[h!]
\begin{center}
\caption{Overview of $\mu_{\rm pe}$ determination from HD$^+$  from the three measurements considered in this article. The second column lists the values of $\mu_{\rm pe}$ derived from the corresponding transition or combination of transitions, with their uncertainties in parentheses. The third column presents the deviations of the extracted values of $\mu_{\rm pe}$ from the CODATA-14 recommended value, $\mu_{\rm pe}$(C14). In the fourth column, the uncertainties of the $\mu_{\rm pe}$ determinations are written in relative terms. Finally in the fifth column, the deviations of the determined values of $\mu_{\rm pe}$ from $\mu_{\rm pe}$(C14) are listed in terms relative to the uncertainty of the determined $\mu_{\rm pe}$.}

\label{muoverview}
\begin{tabular}{ l d d d d }
 \hline
 \hline
 \text{Transition} & \multicolumn{1}{c}{$\mu_{\rm pe}$}  & \multicolumn{1}{c}{$\mu_{\rm pe} $-$\mu_{\rm pe} $(C14)} & \multicolumn{1}{c}{$\delta
   \mu_{\rm pe} $/$\mu_{\rm pe} $ [ppb]} & \multicolumn{1}{c}{($\mu_{\rm pe} $-$\mu_{\rm pe}$(C14))/$\delta \mu_{\rm pe} $} \\
 \hline
 \HDone\                        & 1\,836.152\,689\,2(62)  & 0.000\,015      & 3.4 & 2.5 \\
 \HDfour\                       & 1\,836.152\,684\,6(148) & 0.000\,011      & 8.0 & 0.74 \\
 \HDeight\                      & 1\,836.152\,670\,6(79)  & -0.000\,003\,3  & 4.3 & -0.41 \\
 \text{\HDfour\ and \HDeight\ } & 1\,836.152\,673\,8(70)  & -0.000\,000\,08 & 3.8 & -0.012 \\
 \text{All transitions}         & 1\,836.152\,682\,4(46)  & 0.000\,008\,5   & 2.5 & 1.8 \\
 \hline
 \hline
\end{tabular}
\end{center}
\end{table}

\section{Outlook and conclusion}\label{sec:V}

The uncertainty of $\mu_{\rm pe}$ determined from the three ro-vibrational transitions of HD$^+$ considered in this article is still 27 times larger than the CODATA-14 adjustment of the same. However, the prospects for determination of fundamental constants from spectroscopy of hydrogen molecular ions indicate a substantial possible improvement. In HD$^+$, an experiment towards Doppler-free two-photon spectroscopy of $(v,L):(0,3)\rightarrow(4,2)\rightarrow(9,3)$ is underway in Amsterdam~\cite{Tran2013}. An experimental accuracy better than the theoretical uncertainty of $\sim 1\times 10^{-11}$ would allow a determination of $\mu_{\rm pe}$ (or equivalently, of the proton atomic mass $m_{\rm p}$) with an uncertainty comparable to the CODATA-14 recommended value. Furthermore, $m_{\rm p}$ has recently been measured with a relative uncertainty of 32 parts-per-trillion (ppt) from Penning trap mass measurements~\cite{SturmPRL2017}. However, the measured value of $m_{\rm p}$ is 3~$\sigma$ off from the CODATA-14 value of $m_{\rm p}$. In this context, a precise determination of $m_{\rm p}$ from HD$^+$ may allow a comparison with the $m_{\rm p}$ determination in ref~\cite{SturmPRL2017} and serve as a consistency check for the measurements of the same quantity from different physical systems. A further step would be to combine HD$^+$ and H$_2^+$ spectroscopy at the few-ppt accuracy level in order to constrain not only the $\mu_{\rm pe}$ and $\mu_{\rm dp}$ mass ratios, but also the Rydberg constant $R_{\infty}$ and the nuclear radii $r_{\rm p},r_{\rm d}$, thus shedding light on the curent discrepancies between different determinations of these constants, coloquially known as the proton-radius puzzle~\cite{KarrPRA2016}.

In conclusion, in this article we have revisited the determination of $\mu_{\rm pe}$ from HD$^+$ as done in Ref.~\cite{BiesheuvelNatComm2016,KarshenboimAPB2017} in light of improved theoretical calculations and updated CODATA-14 recommended values of the fundamental constants. The significant improvement in the knowledge of $m_{\rm e}$ from CODATA-10 to CODATA-14 has led to the realization that it is more relevant to use the CODATA value of $\mu_{\rm de}$ in the adjustment, rather than that of $\mu_{\rm dp}$. Hence, we derived the appropriately modified sensitivity coefficients of the transitions concerned and determine the value of $\mu_{\rm pe}$. The thus found value of $\mu_{\rm pe}$ in this work not only differs significantly from the values reported of the same quantity determined from HD$^+$ in ref~\cite{BiesheuvelNatComm2016,KarshenboimAPB2017}, but also possesses a larger error bar. In a similar way we also determine values of $\mu_{\rm pe}$ from other transitions in HD$^+$, based on previously reported experimental results, and we obtain an overall value of $\mu_{\rm pe}$ from all available data in HD$^+$ by a least-squares adjustment. The procedure outlined here could be used for the interpretation of future Doppler-free spectroscopy experiments on HD$^+$.

\section{Acknowledgments}

This work received support from the French-Dutch bilateral EP Nuffic-Van Gogh program. The work of S.P., M.G. and J.C.J.K. was financed by Netherlands Organization for Scientific Research (NWO) through projects 13PR3109, Vidi 12346 and FOM-program `The Mysterious Size of the Proton'. J.-Ph.K. and L.H. acknowledge funding from Agence Nationale de la Recherche (grant ANR-13-IS04-0002-01). J.-Ph.K. acknowledges support from a Fellowship of the Institut Universitaire de France. V.I.K. acknowledges support from the Russian Foundation for Basic Research under Grant No. 15-02-01906-a.


\begin{thebibliography}{99}
\bibitem{KorobovPRL2017} V. I. Korobov, L. Hilico, J.-Ph. Karr, Phys. Rev. Lett.~\textbf{118}, 233001 (2017).
\bibitem{Wing1976} W. H. Wing, G. A. Ruff, W. E. Lamb Jr., J. J. Spezeski, Phys. Rev. Lett.~\textbf{36}, 1488-1491 (1976).
\bibitem{Koelemeij2007} J. C. J. Koelemeij, B. Roth, A. Wicht, I. Ernsting, S. Schiller, Phys. Rev. Lett.~\textbf{98}, 173002 (2007).
\bibitem{Bressel2012} U. Bressel, A. Borodin, J. Shen, M. Hansen, I. Ernsting, S. Schiller, Phys. Rev. Lett.~\textbf{108}, 183003 (2012).
\bibitem{BiesheuvelNatComm2016} J. Biesheuvel, J.-Ph. Karr, L. Hilico, K. S. E. Eikema, W. Ubachs, J. C. J. Koelemeij, Nat. Commun.~\textbf{7}, 10385 (2016).
\bibitem{BiesheuvelAPB2017} J. Biesheuvel, J.-Ph. Karr, L. Hilico, K. S. E. Eikema, W. Ubachs, J. C. J. Koelemeij, Appl. Phys. B \textbf{123}: 23 (2017).
\bibitem{KarshenboimAPB2017} S. G. Karshenboim, V. G. Ivanov, Appl. Phys. B \textbf{123}:18 (2017).
\bibitem{Schil05} S. Schiller, V. Korobov, Phys. Rev. A~\textbf{71}, 032505 (2005).
\bibitem{Karr06} J.-Ph. Karr, L. Hilico, J. Phys. B~\textbf{39}, 2095 (2006).
\bibitem{Li07} H. Li, J. Wu, B.-L. Zhou, J.-M. Zhu, Z.-C. Yan, Phys. Rev. A~\textbf{75}, 012504 (2007).
\bibitem{Ning14} Y. Ning, Z.-C. Yan, Phys. Rev. A~\textbf{90}, 032516 (2014).
\bibitem{Kor06} V. I. Korobov, Phys. Rev. A~\textbf{74}, 052506 (2006).
\bibitem{Kor08} V. I. Korobov, Phys. Rev. A~\textbf{77}, 022509 (2008).
\bibitem{Kor14PRA} V. I. Korobov, L. Hilico, J.-Ph. Karr, Phys. Rev. A~\textbf{89}, 032511 (2014).
\bibitem{Kor14PRL} V. I. Korobov, L. Hilico, J.-Ph. Karr, Phys. Rev. Lett.~\textbf{112}, 103003 (2014).
\bibitem{Kor16} V. I. Korobov, J. C. J. Koelemeij, L. Hilico, J.-Ph. Karr, Phys. Rev. Lett.~\textbf{116}, 053003 (2016).
\bibitem{Karr17} J.-Ph. Karr, L. Hilico, V. I. Korobov, Phys. Rev. A~\textbf{95}, 042514 (2017).
\bibitem{Zhong12} Z.-X. Zhong, P.-P. Zhang, Z.-C. Yan, T.-Y. Shi, Phys. Rev. A~\textbf{86}, 064502 (2012).
\bibitem{Kor12} V. I. Korobov, Z.-X. Zhong, Phys. Rev. A~\textbf{86}, 044501 (2012).
\bibitem{Koelemeij2012} J. C. J. Koelemeij, D. W. E. Noom, D. de Jong, M. A. Haddad, W. Ubachs, Appl. Phys. B  \textbf{107}: 1075 (2012).
\bibitem{KarrPRA2016} J.-Ph. Karr, L. Hilico, J. C. J. Koelemeij, V.I. Korobov, Phys. Rev. A~\textbf{94}, 050501 (2016).
\bibitem{CODATA10} P. J. Mohr, B. N. Taylor, D. B. Newell, Rev. Mod. Phys.~\textbf{84}, 1527 (2012).
\bibitem{CODATA14} P. J. Mohr, D. B. Newell, B. N. Taylor, Rev. Mod. Phys.~\textbf{88}, 035009 (2016).
\bibitem{Tran2013} V. Q. Tran, J.-{\relax Ph}. Karr, A. Douillet, J. C. J. Koelemeij, L. Hilico, Phys. Rev. A \textbf{88}, 033421 (2013).
\bibitem{SturmPRL2017} F. Hei{\ss}e, F. K{\"o}hler-Langes, S. Rau, J. Hou, S. Junck, A. Kracke, A. Mooser, W. Quint, S. Ulmer, G. Werth, K. Blaum, S. Sturm, Phys. Rev. Lett.~\textbf{119}, 033001 (2017).


\end{thebibliography}
\end{document}